\documentstyle[prl,aps,floats,epsf]{revtex}

\begin{document}
            
\draft

\twocolumn[\hsize\textwidth\columnwidth\hsize\csname @twocolumnfalse\endcsname
\title{Site-Dilution-Induced Antiferromagnetic Long-Range Order
in Two-Dimensional Spin-Gapped Heisenberg Antiferromagnet}

\author{Chitoshi~Yasuda, Synge~Todo\cite{todo}, Munehisa~Matsumoto, and Hajime~Takayama} 

\address{Institute for Solid State Physics, University of Tokyo, Kashiwa 277-8581, Japan}

\date{\today}

\maketitle

\widetext

\begin{abstract}
 Effects of the site dilution on spin-gapped Heisenberg antiferromagnets
 with $S=1/2$ and $S=1$ on a square lattice are investigated by means of
 the quantum Monte Carlo method. It is found that effective magnetic
 moments induced around the diluted sites exhibit the antiferromagnetic
 long-range order in the medium of spin-singlet pairs. Their microscopic
 structure is examined in detail and important roles of the higher
 dimensionality than one on the phenomenon are discussed.
\end{abstract}

]

\narrowtext

The nonmagnetic-impurity-induced, or more specifically, the
site-dilution-induced antiferromagnetic long-range order (AF-LRO) in
quasi-one-dimensional (Q1D) quantum antiferromagnets has attracted
many researchers more than a decade~\cite{shender}. Typically it has been
observed in the first inorganic spin-Peierls (SP) compound
CuGeO$_3$~\cite{hase} by the substitution of nonmagnetic Zn for
magnetic Cu with $S=1/2$ spin. In the induced AF state with sufficiently
weak dilution the lattice dimerization is preserved~\cite{martin}. This
strongly suggests that at least this site-dilution-induced transition is
dominantly of a magnetic origin, though for thorough understanding of the
SP transition in the system other ingredients such as the
magneto-elastic effect~\cite{fukuyama} is indispensable. Another example
of the site-dilution-induced AF-LRO has been observed in a Q1D Haldane
compound Pb(Ni$_{1-x}$Mg$_x$)$_2$V$_2$O$_8$ where Ni carries $S=1$
spin~\cite{uchiyama}.
 
A most plausible scenario for the site-dilution-induced AF-LRO so far
discussed is as follows~\cite{shender,nagaosa,martins,imada,wessel}. The
mother compound ($x=0$) has a spin-gapped ground state such as the dimer
singlet state in CuGeO$_3$ and the Haldane state in
PbNi$_2$V$_2$O$_8$. When $x>0$, spins on the sites neighboring the
diluted sites organize a spin cluster which is a close resemblance of
the edge state in the Haldane chain~\cite{miyashita}. Here we call it as
a whole an `{\it  effective spin}'. Between the effective spins there
exist the effective interactions mediated by spins which are in the
singlet state. The averaged magnitude of the effective interactions is
exponentially small as a function of the concentration of nonmagnetic
impurities. Although
they are either ferromagnetic or AF depending on relative positions of
the effective spins, the staggered nature with respect to the original
lattice is completely preserved, and so the effective spins are
antiferromagnetically ordered at $T=0$ in more than two dimensions.
The very last part of the above scenario, i.e., the effect of the higher
dimensionality, has been mostly treated by the mean-filed-like arguments
except for a few works~\cite{imada,wessel}.

In the present work we have extensively performed the quantum Monte
Carlo (QMC) analysis on the site-diluted 2D Heisenberg antiferromagnets
on a square lattice, and have reached the same scenario, but with
evidences which support it more directly than the previous works. One of
 them is the microscopic spatial information on the effective spins
induced by the dilution. We also present the result consistent with the
scenario in the 2D coupled Haldane chain system with $S=1$.

The quantum spin system of our present interest is described by the
Hamiltonian
\begin{eqnarray}
   \label{ham}
   {\cal H}&=&\sum_{i,j}\epsilon_{2i,j}\epsilon_{2i+1,j}
             {\bf S}_{2i,j}\cdot{\bf S}_{2i+1,j} \nonumber \\
    &+&\alpha\sum_{i,j}\epsilon_{2i+1,j}\epsilon_{2i+2,j}
             {\bf S}_{2i+1,j}\cdot{\bf S}_{2i+2,j} \\
    &+&J'\sum_{i,j}\epsilon_{i,j}\epsilon_{i,j+1}
             {\bf S}_{i,j}\cdot{\bf S}_{i,j+1} \ , \nonumber
\end{eqnarray}
where 1 and $\alpha$ ($0 \leq \alpha \leq 1$) are the intrachain
alternating coupling constants,  $J'$ ($\geq 0$) is the interchain AF
coupling constant and ${\bf S}_{i,j}$ is the quantum spin operator at site
$(i,j)$. Randomly quenched magnetic occupation factors $\{\epsilon_{i,j}\}$
independently take either 1 or 0 with probability $1-x$ and $x$,
respectively, where $x$ is the concentration of the diluted sites.
 
In the present work the QMC simulations with the
continuous-imaginary-time loop algorithm~\cite{MC} is carried out on the
$S=1/2$ and $S=1$ systems on $L\times L$ square lattices with the periodic
boundary condition. For diluted systems with $x>0$ the number of samples
averaged over is $100 \sim 700$ depending on $x$. For each samples,
$10^4$ Monte Carlo steps (MCS) are spent for measurement after $10^3$
MCS for thermalization. The staggered magnetization, $M_{\rm s}(x)$, at
zero temperature is evaluated by
\begin{equation}
   M_{\rm s}^{2}(x)=\lim_{L\to \infty}\lim_{T\to 0}
       \frac{3S_{\rm s}(L,T,x)}{L^{2}} \ ,
\end{equation}
where
\begin{equation}
    \label{str}
    S_{\rm s}(L,T,x) \equiv \frac{1}{L^{2}}\sum_{i,j}
        (-1)^{|r_{i}-r_{j}|}
        \langle S_{i}^{z}S_{j}^{z} \rangle
\end{equation}
is the static staggered structure factor. The bracket
$\langle \cdots \rangle$ in Eq.(\ref{str}) denotes both the thermal and
random averages.  The value of $S_{\rm s}(L,T,x)$ converges to its
zero-temperature value at $T$ lower than either a gap due to the
finiteness of the system or the intrinsic spin gap. Thus $S_{\rm s}(L,T,x)$
at low temperatures where its $T$-dependence becomes not discernible within 
the error bars is taken as an estimate of $S_{\rm s}(L,0,x)$.
 
\begin{figure}[t]
 \epsfxsize=0.47\textwidth
 \epsfbox{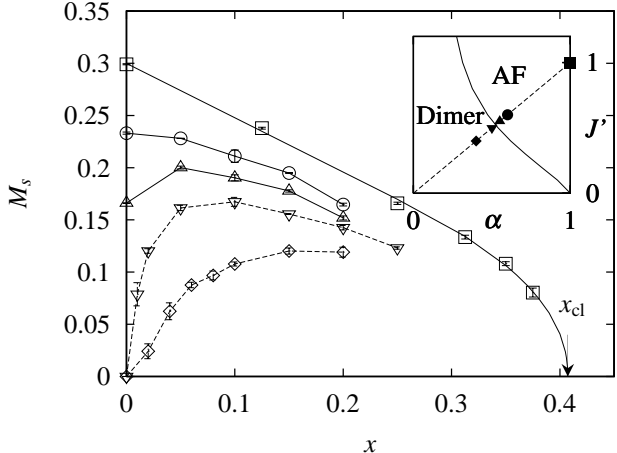}
 \caption{Concentration dependence of the staggered magnetization at
 zero temperature for systems with $\alpha=J'=1$, 0.6, 0.55, 0.5, and
 0.4. The $\alpha-J'$ phase diagram of the pure systems is shown in the
 inset. The data for $\alpha=J'=1$ are results in ref. [13] with 
 $x_{\rm cl} (\simeq 0.407254)$ being the percolation threshold. All the
 lines are guides to eyes. }
\end{figure}

First we determine the ground-state phase diagram of the pure systems on
the $\alpha - J'$ plane. Its details will be discussed
elsewhere~\cite{matsumoto}. Here we simply show the result for the $S=1/2$
system in the inset of Fig. 1, and mention that, except for the critical
point $(\alpha, J')=(1,0)$, the quantum dimer-AF transition we obtain is
ascertained to belong to the same universality class as that of the 3D
classical Heisenberg model.

Now let us concentrate on the site-dilution-induced AF-LRO in the
$S=1/2$ systems with $\alpha=J'$ indicated in the inset of
Fig. 1. Expecting that the fundamental aspects of the phenomenon do not
crucially depend on the values of $\alpha$ and $J'$, we choose
those on the line $\alpha=J'$. It connects the system with 
$\alpha=J'=0$, which consists of a set of independent dimers, and the 
isotropic antiferromagnet with $\alpha=J'=1$. Along this line, the
quantum dimer-AF phase transition occurs at $\alpha=\alpha_{\rm
c}=0.52337(3)$~\cite{matsumoto}. The dilution effects on the
isotoropic system was investigated in details in the previous 
works~\cite{kato,yasuda}; as shown in Fig. 1, the site-dilution simply 
reduces $M_{\rm s}$ which exists already in the pure system. For the 
systems with $\alpha=J'=0.5$ and $0.4$, on the other hand, $M_{\rm s}$, 
which is zero for $x=0$, becomes finite even with 1 or 2 $\%$ of site 
dilution. This is nothing but the site-dilution-induced AF-LRO of our 
present interest.

\begin{figure}[t]
 \epsfxsize=0.47\textwidth
 \epsfbox{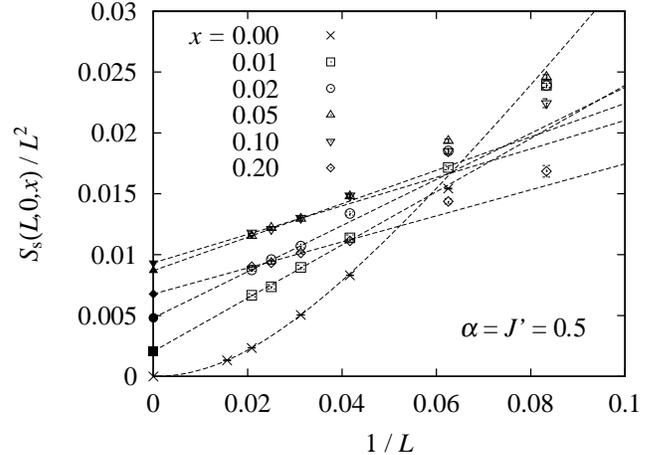}
 \caption{System-size dependence of $S_{\rm s}(L,0,x)/L^{2}$ in the case of
 $\alpha=J'=0.5$. Dashed lines are obtained by the least-squares fitting
 for the largest three system sizes for each $x$. The extrapolated
 values are denoted by solid symbols. }
\end{figure}

The values of $M_{\rm s}$ in Fig. 1 are extracted from the $L$-dependence
of $S_{\rm s}(L,0,x)/L^{2}$ as shown in Fig. 2. The data for $x=0$ are well
fitted to $S_{\rm s}(L,0,x)/L^{2}\propto (1-{\rm exp}(-L/\xi_{\rm p}))/L^{2}$,
where $\xi_{\rm p}$ is the correlation length. This form is derived by
the modified spin-wave theory for a spin-gapped state. For diluted systems
with $x>0$, on the other hand, $M_{\rm s}$ is evaluated by fitting the
data to $S_{\rm s}(L,0,x)/L^{2}\simeq M_{\rm s}^{2}/3+a/L$ which is
obtained, for example, by the linear spin-wave theory on the state with a
finite $M_{\rm s}$.

The above-mentioned correlation length of the pure system, $\xi_{\rm p}$,
is given by $\xi_{\rm p}=(\xi_{\rm p}^{x}\xi_{\rm p}^{y})^{1/2}$ where
$\xi_{\rm p}^{x}$ and $\xi_{\rm p}^{y}$ are the anisotropic correlation
lengths. They are estimated from the dynamic staggered correlation function
by using the second moment method. Similarly the spin gap, 
$\Delta_{\rm p}$, is evaluated from behavior of the function on the
imaginary-time axis. We obtain~\cite{matsumoto} $\xi_{\rm p}^{x}=3.0089(9)$, 
$\xi_{\rm p}^{y}=2.2097(6)$ and $\Delta_{\rm p}=0.32255(3)$ for 
$\alpha=J'=0.4$ and $\xi_{\rm p}^{x}=11.998(9)$,
$\xi_{\rm p}^{y}=9.312(10)$ and $\Delta_{\rm p}=0.0918(1)$ for
$\alpha=J'=0.5$. The nearer
to the transition point a system is, the smaller is $\Delta_{\rm p}$ and
the larger is $\xi_{\rm p}$. 

In order to gain insights into nature of the site-dilution-induced AF-LRO
obtained above, we investigate the local static staggered structure
factor defined by
\begin{equation}
    \label{chi}
    {\cal S}(i) \equiv
       \sum_{j}(-1)^{|r_{i}-r_{j}|}\langle S_{i}^{z}S_{j}^{z} \rangle \ .
\end{equation}
Figure 3 shows the real-space distribution of ${\cal S}(i)$ for 
$\alpha=J'=0.5$ on a $64\times 64$ lattice from which 30 spins are 
randomly removed. The temperature, $T=0.001$, can be regarded as zero
temperature for the system with these parameters. The points protruding
downwards to zero are on the diluted sites. The peaks of ${\cal S}(i)$
appear on the sites connected by a strong bond 1 with these diluted
sites. Naturally they are considered to be the contribution of the spins
which get rid of the singlet pairing by the dilution. The peaks exhibit
a significant spatial extent whose size is given by $\xi_{\rm p}^{x}$
and $\xi_{\rm p}^{y}$ evaluated above.

\begin{figure}[t]
 \epsfxsize=0.5\textwidth
 \epsfbox{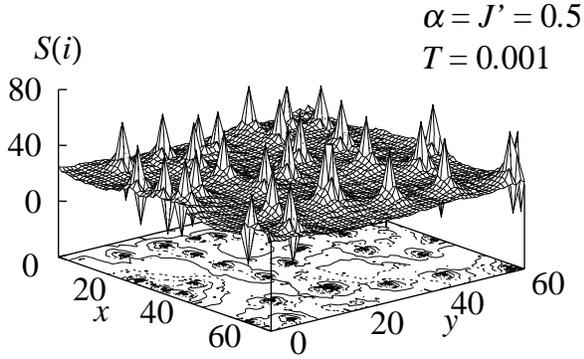}
 \caption{3D plot of the local static staggered structure factor in the fixed
 configuration of 30 impurities on a 64$\times$64 lattice for
 $\alpha=J'=0.5$ at $T=0.001$. 
 Contours are shown in the bottom.}
\end{figure}

\begin{figure}[t]
 \epsfxsize=0.5\textwidth
 \epsfbox{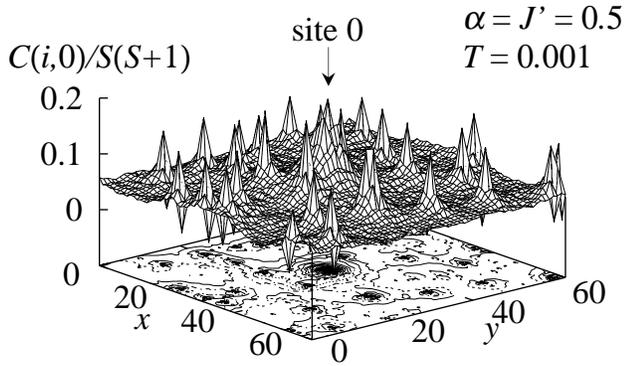}
 \caption{3D plot of the staggered correlation function from the peak of
 an induced magnetic moment on site 0 (=(31,31)) for the same dilution
 configuration as Fig. 3. 
 } 
\end{figure}

From the analogy of the 1D spin-gapped state~\cite{nagaosa}, the exchange
coupling between the two effective spins centered at sites $m$ and $n$ is
expected to be given by
${\tilde J}_{mn} \propto (-1)^{|r_{m}-r_{n}+1|}
{\rm exp}[-l(\xi_{\rm p}^{x}\xi_{\rm p}^{y})^{-1/2}]$, where
$l=|r_{m}-r_{n}|$ is the distance between the effective spins. For a
sufficiently weak dilution, we can naturally expect that the effective
Hamiltonian, ${\cal H}_{\rm eff}=
\sum_{<mn>}{\tilde J}_{mn}\tilde{{\bf S}}_{m}\cdot \tilde{{\bf S}}_{n}$
well describes magnetic behavior associated with the
site-dilution-induced AF-LRO. As we have already pointed out, 
${\tilde J}_{mn}$ is considered to completely preserve the staggered
nature with respect to the original lattice since the original
interactions are all AF with no frustration at all. This is in fact
confirmed by Fig. 4 where we draw the static staggered correlation
function  $C(i,0)=(-1)^{|r_{i}-r_{0}|}\langle S_{i}^{z}S_{0}^{z}\rangle$
with site $0\ (=(31,31))$ being the peak position of an effective spin.
The system is the same as that in Fig. 3. At any site $i$, $C(i,0)$ is
positive except for the diluted sites where it is zero. It peaks at
sites of other effective spins with heights almost independent of the
distance from site 0, while it stays at a relatively lower value on the
sites far from the diluted sites where spins form the singlet state.
Thus Fig. 4 combined with Fig. 3 clearly demonstrates that the
site-dilution-induced AF-LRO is mostly carried by the effective spins.
Each of them contributes to $M_{\rm s}$ by the amount proportional to
$\xi_{\rm p}^2$. This explains a sharper rise near $x\simeq 0$ of
$M_{\rm s}$ of $\alpha=J'=0.5$ than that of $\alpha=J'=0.4$ in Fig. 1.
It contributes to the uniform susceptibility, on the other hand, by the
amount $|\tilde{{\bf S}}_m| = 1/2$~\cite{nagaosa} as discussed below.

The averaged distance $\langle l \rangle$ between diluted sites is
proportional to $1/\sqrt{x}$. The typical coupling which scales the
excitation energy
of the spin-wave-like modes is then given by $\langle |{\tilde J}_{mn}|
\rangle \propto
    {\rm exp}(-1/\sqrt{x\xi_{\rm p}^{x}\xi_{\rm p}^{y}})$.
It becomes much smaller than $\Delta_{\rm p}$ for a sufficiently small
value of $x$. Thus we can naturally expect the existence of low-lying
excitation modes well separated from the original triplet excited state
with $\Delta_{\rm p}$ as has been in
fact observed by the neutron inelastic experiment in the doped
CuGeO$_3$~\cite{martin,masuda,regnault2}. Another peculiar feature of
the present site-dilution-induced AF-LRO is that it is inhomogeneous in
a shorter length scale than $\langle l \rangle$. This may explain the
experimental fact~\cite{martin} that the spectrum of the spin-wave-like
mode becomes broader for the larger wave number.

\begin{figure}[t]
 \epsfxsize=0.47\textwidth
 \epsfbox{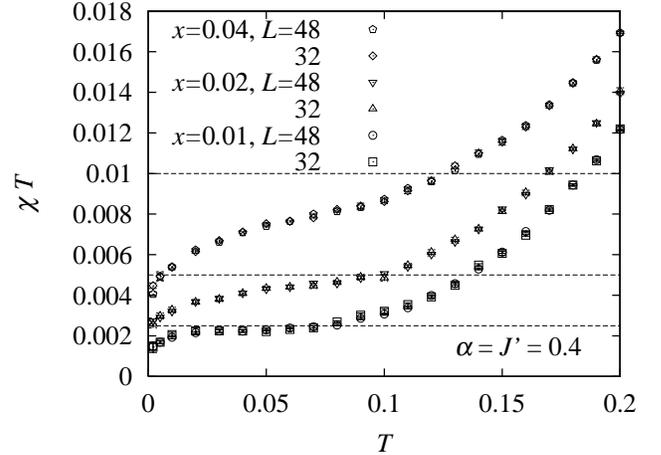}
 \caption{Temperature dependence of the effective Curie constant for
 $\alpha=J'=0.4$. The broken lines show $\chi T=x/4$ for $x=0.04$, 0.02,
 and  0.01 from top.}
\end{figure}

By the present simulation the existence of low-lying excitation modes
mentioned above is ascertained by the $T$-dependence of the effective
Curie constant $\chi T$~\cite{wessel} which is shown in Fig. 5 for
$\alpha=J'=0.4$. For $x=0.01$ the plateau of a height nearly equal to
$\chi T = x/4$ in unit of $ g^{2}\mu_{\rm B}^{2}/k_{\rm B}$ is clearly
seen around $T \sim 0.05$ which is smaller than $\Delta_{\rm p}$ by an
order of magnitude. We can attribute the plateau to the Curie law of the
effective spins with $|\tilde{{\bf S}}_m| = 1/2$ as mentioned above.
They start to correlate at lower temperatures than 
$\langle |{\tilde J}_{mn}| \rangle$. Already for $x=0.02$, however, the
plateau is obscured, indicating that $\langle |{\tilde J}_{mn}| \rangle$
becomes comparable in magnitude with $\Delta_{\rm p}$. Qualitatively
similar features have been observed experimentally~\cite{manabe}.

Essentially the same scenario as the $S=1/2$ case is expected to hold also
for similar antiferromagnets with $S=1$. As their representative we have
examined the coupled Haldane chain system with $\alpha=1$ and $J'=0.04$ in
Eq.(\ref{ham}). The latter is a little smaller than the critical value
$J'_{\rm c}=0.04365(1)$ for the Haldane-AF transition
point~\cite{matsumoto}. Behavior of its $M_{\rm s}(x)$ shown in Fig. 6 is
quite similar to that in Fig. 1, and further confirms the above-mentioned
scenario for the site-dilution-induced AF-LRO. 

Here it is worth noting that the interchain interaction $J'$ introduces 
another energy scale than 
$\Delta_{\rm p}$ and $\langle |{\tilde J}_{mn}| \rangle$ into the
problem of the present interest. It is the coupling strength, ${\cal J}$,
of spins on the two opposite neighbors of the diluted site through the
shortest interaction paths. It is {\it ferromagnetic} and of the order of
$J'^2$. For a sufficiently small $x$,
$\langle |{\tilde J}_{mn}| \rangle \ll {\cal J} \ll \Delta_{\rm p}$
holds. In such a coupled Haldane chain system the
following successive process is expected to occur; effective
spins with $S=1/2$, instead of $S=1$, induced at both sides of diluted
sites are independently fluctuating at ${\cal J} \ll T \ll \Delta_{\rm p}$,
the two spins are coupled to behave as one effective spin of $S=1$ at 
$\langle |{\tilde J}_{mn}| \rangle \ll T \ll {\cal J}$, and then the 
$S=1$ effective spins begin to correlate antiferromagnetically at 
$T \ll \langle |{\tilde J}_{mn}| \rangle$. For the Haldane chain system 
in Fig. 6, $\Delta_{\rm p} \gg {\cal J} \sim 0.0016$ is satisfied. 
Unfortunately, however, the expected two-plateau structure in $\chi T$
is hard to be seen, since ${\cal J}$ is too small. In the $S=1/2$ system
of Fig. 1, on the other hand, ${\cal J} \sim \Delta_{\rm p}$, and there
appears one plateau at $T \ll {\cal J}$ as already shown in Fig. 5.

\begin{figure}[t]
 \epsfxsize=0.47\textwidth
 \epsfbox{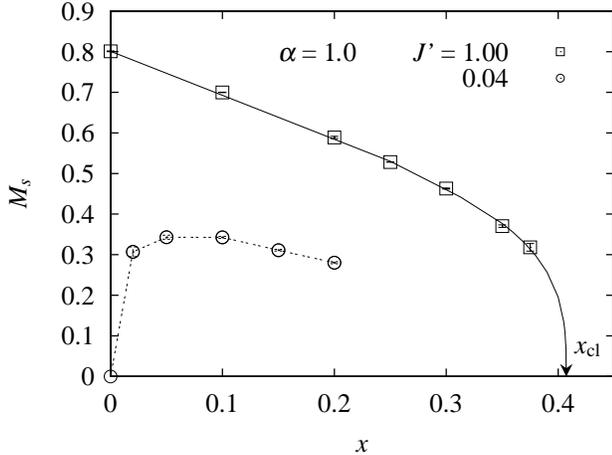}
 \caption{The staggered magnetization of the coupled
 Haldane chain systems with $J'=1$ and 0.04. The data for $\alpha=J'=1$
 are results in ref. [13]. All the lines are guides to eyes.}
\end{figure}

Another interesting problem on which ${\cal J}$ plays an essential role
is a possible difference between the AF-LRO's induced by site-dilution
and bond-dilution. In the latter case the coupling ${\cal J}$ between
the two end spins of a diluted bond becomes {\it AF}.  Therefore,
if $|{\cal J}|\ (\sim J'^2)$ is significantly larger than
$\langle |{\tilde J}_{mn}| \rangle$, the two spins may
again form a singlet pair and cannot contribute to the AF-LRO. This
implies a possible existence of a critical concentration $x_{\rm c}$
below which the AF-LRO does not come out. In contrast, there is no 
reason for a finite $x_{\rm c}$ in the site-diluted case discussed
above. Experimentally, however, no drastic difference has
been observed between site-diluted and bond-disordered
CuGeO$_3$ yet~\cite{regnault2}. For quantitative understanding of the
experimental resuts we have to take into account other ingredients than
those in Eq.(\ref{ham}) such as the the next-nearest-neighboring
intrachain interactions~\cite{castilla}. This is beyond the scope of the
present work. 

In summary, we have argued based on the QMC simulation on the
2D AF Heisenberg models that dilution of spins in a system having the
spin-gapped ground state induces effective spins which are strongly
correlated. As a consequence there occurs the AF-LRO state which has
low-lying  excitation states well separated from the original triplet
excited state. 

Most of numerical calculations in the present work have been performed
on the SGI 2800 at Institute for Solid State Physics, University of
Tokyo. The present work is supported by the ``Research for the Future
Program'' (JSPS-RFTF97P01103) of Japan Society for the Promotion of
Science.


\begin{thebibliography}{99}

\bibitem[*]{todo} Present address: Theoretische Physik,
	   Eidgen\"ossische Technische Hochschule, CH-8093 Z\"urich,
	   Switzerland.
\bibitem{shender} E. F. Shender and S. A. Kivelson, Phys. Rev. Lett. 
{\bf 66}, 2384 (1991).
\bibitem{hase} M. Hase, I. Terasaki and K. Uchinokura, Phys. Rev. Lett. 
{\bf 70}, 3651 (1993).
\bibitem{martin} M. C. Martin {\it et al}., Phys. Rev. B {\bf 56}, 3173
	(1997).
\bibitem{fukuyama} H. Fukuyama, T. Tanomoto and M. Saito, J. Phys. Soc. Jpn. 
{\bf 65}, 1182 (1996). 
\bibitem{uchiyama} Y. Uchiyama {\it et al}., Phys. Rev. Lett. {\bf 83},
	   632 (1999).
\bibitem{nagaosa} N. Nagaosa {\it et al}., J. Phys. Soc. Jpn. {\bf 65},
	3724 (1996); Y. Iino and M. Imada, J. Phys. Soc. Jpn. {\bf 65}, 3728
	   (1996). 
\bibitem{martins} G. B. Martins {\it et al}., Phys. Rev. Lett. {\bf 78}, 3563 
(1997).
\bibitem{imada} M. Imada and Y. Iino, J. Phys. Soc. Jpn. {\bf 66}, 568
	   (1997).
\bibitem{wessel} S. Wessel {\it et al}., Phys. Rev. Lett. {\bf 86}, 1086 (2001). 
\bibitem{miyashita} S. Miyashita and S. Yamamoto, Phys. Rev. B {\bf 48},
	913 (1993).
\bibitem{MC} H. G. Evertz, G. Lana and M. Marcu, Phys. Rev. Lett. {\bf 70}, 875
	(1993); B. B. Beard and U.-J. Wiese, Phys. Rev. Lett. {\bf 77}, 
5130 (1996); S. Todo and K. Kato, cond-mat/9911047 (unpublished).
\bibitem{matsumoto} M. Matsumoto {\it et al}., (unpublished).
\bibitem{kato} K. Kato {\it et al}., Phys. Rev. Lett. {\bf 84}, 4204
	(2000). 
\bibitem{yasuda} C. Yasuda {\it et al}., Phys. Rev. B {\bf 63},
	   140415(R) (2001).
\bibitem{masuda} T. Masuda {\it et al}., Phys. Rev. Lett. {\bf 80}, 4566 
(1998).
\bibitem{regnault2} L. P. Regnault {\it et al}., Europhys. Lett. {\bf 32}, 
579 (1995). 
\bibitem{manabe} K. Manabe {\it et al}., Phys. Rev. B {\bf 58}, R575 (1998).
\bibitem{castilla} G. Castilla, S. Chakravarty and V. J. Emery,
	   Phys. Rev. Lett. {\bf 75}, 1823 (1995).

\end{thebibliography}
\end{document}